\documentclass[aps,prb,twocolumn,showpacs,epsfig]{revtex4}
\usepackage{graphicx} 
\begin{document}
\title
{Interrupted chain assisted Al atomic wires on Si(211)}

\author{Bikash C Gupta and  Inder P. Batra}
\affiliation{ Department of Physics, 845 W Taylor street,
University of Illinois at Chicago, Chicago, Illinois 60607-7059,
USA}

\date{\today}

\begin{abstract}
Possibility for the formation of stable Al atomic wire on the
Si(211) surface is investigated using density functional theory
based total energy calculations. The stable adsorption sites and the
surface structures at various sub-monolayer coverages of Al are
presented. It is found that the most stable and natural surface
structures around one monolayer coverage is either $1\times5$ or
$1\times6$ which agrees with experimental observations. More
significantly, our study revealed that unlike the case of Ga the
formation of continuous atomic Al chains assisted by interrupted Al
chains may be possible by controlling the experimental conditions
(pressure and temperature) such that the $\mu_{\rm Al}$ remains
within $\sim 0.3$ eV from the bulk value. While the Al covered
Si(211) surface may be metallic or semiconducting, the Ga covered
Si(211) is always semiconducting in nature below 1 ML coverage. Also
significant is the fact that Al atoms move from groove sites to the
sites between terrace and trench atoms through lower energy channels
as the coverage goes from 1/8 monolayer to 1/4 monolayer.
\end{abstract}


\maketitle

\section{Introduction}

Technological importance of nanowire structures on silicon
substrates has generated much interest in studying structural
arrangements of metals on various Si substrates. \cite{him1,him2} In
particular, metals like Al, Ga and In at sub-monolayer coverages on
the Si(001) surface have been examined \cite{ipb1,ipb2} for the
formation of stable atomic wires on surfaces. Though, stable atomic
wire structures have not been found on the bare Si(001) surface,
there has been some progress \cite{wat1,wat2,ipb3} towards the
formation of stable atomic wires on the patterned hydrogen
terminated Si(001) surface.

Recently, the vicinal Si surfaces have attracted much attention for
the growth of self assembled atomic wires on them. The Si(211)
surface \cite{chad,kapl} is one of the vicinal Si surfaces that is
being widely used by experimentalists \cite{yate,glem,bas1,gonz} as
a substrate for growing atomic wire structures. This surface can be
viewed as a stepped arrangement of narrow (111) terraces.

\begin{figure}
  \includegraphics[width=3in]{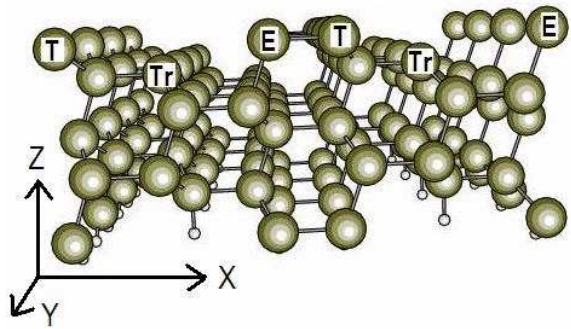}\\
  \caption{Atoms in perspective form the ideal Si(211) $2 \times 4$
  super-cell. The bottom layer Si atoms are passivated by hydrogen
  atoms (small circles). Surface terrace, trench and edge atoms are
  denoted as T, Tr and E respectively. X, Y and Z directions
  correspond to [$\bar{1} 1 1$], [$0 1 \bar{1}$] and [$2 1 1$]
  respectively }
  \label{fig1}
\end{figure}

A three dimensional view of a small portion of the ideal Si(211)
surface is shown in Fig.~\ref{fig1}. The atoms marked T (called the
terrace atoms) on the terrace are three-fold coordinated and thus
have one dangling bond each; those on the step edge, marked E
(called the edge atoms) are two-fold coordinated and have two
dangling bonds each. Second layer Si atoms which are not directly
bonded either with terrace or edge atoms are denoted as Tr (called
the trench atoms) have one dangling bond each. The Si(211) surface
consists of two-atom wide terraces between terrace and edge atoms
along the [$\bar{1}11$] direction. Two consecutive terraces are
separated by steps and are 9.4 \AA~ apart in the [$\bar{1}11$]
direction, while they extend infinitely along [$01\bar{1}$]. The
reason for using Si(211) is that the metals are expected to nucleate
at the step edges, and form atomic nanowires.

The scanning tunneling microscopy (STM) images obtained by Baski et
al. \cite{bas1} for the Ga/Si(211) system showed the formation of
atomic wires of Ga extending along the [$01\bar{1}$] direction with
vacancies at every fifth or sixth atoms interval {\em i.e.}, the
surface structure becomes $1 \times$N with N=5-6. The Ga wires were
separated by 9.4 \AA~ along the [$\bar{1}11$] direction. These
authors also provided a theoretical model supporting their
experimental observations. The same system was also studied by
Gonzalez et al. \cite{gonz} and their STM images revealed the
formation of two atomic chains of Ga extending along the
[$01\bar{1}$] direction within the 9.4 \AA~ distance along the
[$\bar{1}11$] direction. They also found that both the chains have
vacancies at six atoms interval along the [$01\bar{1}$] direction
{\em i.e.}, the surface structure is $1 \times 6$. The authors, in
addition, performed a detailed theoretical calculations in support
of their experimental results. From the experimental and theoretical
results for Ga/Si(211), it is clear that a stable uninterrupted Ga
chain can not be formed on the Si(211) substrate. The In/Si(211)
system has also been studied experimentally by means of STM, low
energy electron diffraction (LEED) and Auger electron spectroscopy
(AES). It is found that the surface pattern is $1 \times 7$.
\cite{zgai} They also presented an intuitive model in support of
their results. Wang et al. \cite{wang} performed LEED experiments on
the Al/Si(211) at sub-monolayer coverage and they found the surface
pattern to be $1 \times 6$. The question we raise here is whether or
not a stable uninterrupted atomic Al chain may be formed on the
Si(211) substrate. Since, no detailed electronic structure
calculations have so far been made for the precise atomic
arrangements of Al on the Si(211) surface, the possibility of
formation of a stable continuous Al wire on Si(211) is also worth
investigating. In this paper we therefore perform extensive total
energy calculations for the Al/Si(211) system.

From theoretical perspective it is important to know the
reconstruction of the clean surface prior to the placement of any
metal on the surface. As far as the reconstruction of the clean
Si(211) surface is concerned, there have been several studies
\cite{wang,berg,sen,bas2}. However, Baski et al. \cite{bas3} found
(using STM) that the clean Si(211) is really unstable and consists
of nanofacets with (111) and (337) orientations. It has also been
shown that the Si(211) orientation is regained upon metal
adsorption.\cite{bas1,gonz} In view of this result, \cite{bas1,gonz}
we therefore consider the ideal bulk terminated Si(211) substrate
for the Al atomic wires study. In our investigations, we
systematically increase Al coverage on the Si(211) surface and see
how the adsorption sites and the surface structure change as we go
towards a monolayer (ML) coverage. We will show that atomic
arrangement of Al on Si(211) will depend on the chemical potential
of Al ($\mu_{{\rm Al}}$) which in turn depends on the experimental
conditions such as the substrate temperature, the Al vapor pressure
in the effusion cell and the Al vapor pressure on the substrate. Our
study reveals that the interrupted Al chains with period five/six
(similar to the case of Ga) may be formed. However, most significant
result in this investigation is that a stable uninterrupted Al chain
extending along the [$01\bar{1}$] direction assisted by an
interrupted chain can be formed at about 1 ML coverage.

The paper is organized as follows. In Section~\ref{sec:method} we
present the parameters used in the pseudopotential density
functional calculations. The results and discussions are presented
in Section~\ref{sec:results}. Finally, in section~
\ref{sec:summary}, we summarize our principal findings.

\section{Method}
\label{sec:method} Total energy minimization calculations are
carried out within the density functional theory (DFT) in
conjunction with the pseudopotential approximation. The Si(211)
surface is represented in a repeated slab geometry. Each slab
contains seven Si(211) layers with a vacuum region of 12 \AA. For
example, in the $1 \times 4$ super-cell, each layer contains eight
Si atoms: two along [$\bar{1} 1 1$] and four along [$0 \bar{1} 1$].
The top layer (within the 1$\times$4 super-cell) contains four edge
and four terrace Si atoms. It is noted that in an ideal Si(211):$1
\times 1$ surface each layer consists of two distinct atoms. The Si
atoms in the bottom layers have their dangling bonds saturated by
hydrogen atoms (see Fig.~\ref{fig1}). Since the edge atoms have two
dangling bonds each, the trench atoms have one dangling bond each
and the terrace atoms have one dangling bond each, we require
sixteen hydrogen atoms to saturate all the dangling bonds at the
bottom of the slab in the $1 \times 4$ super-cell. We have used
super-cells of various sizes ($1 \times 2$, $1 \times 3$, $1 \times
4$, $1 \times 5$, $1 \times 6$, and $1 \times 7$) in our
calculations. The top five Si layers are relaxed for geometry
optimization while the two lower-most Si layers and the hydrogen
atoms are held fixed to simulate the bulk-like termination. The wave
functions are expanded in a plane wave basis set with a cutoff
energy $|\vec{k} + \vec{G}|^2 \le 250$ eV. The Brillouin zone (BZ)
integration is performed within a Monkhorst-Pack (MP) \cite{mank}
scheme using four inequivalent k-points. It has been established
earlier \cite{sen} that the energy cutoff, the set of k-points, the
number of layers in the slab and the amount of vacuum region
considered here give sufficiently converged values for total
energies. Ionic potentials are represented by Vanderbilt-type
ultra-soft pseudopotentials~\cite{ultr} and results are obtained
using generalized gradient approximation (GGA)~\cite{pw91} for the
exchange-correlation potential. Preconditioned conjugate gradient is
used for wave function optimization and a conjugate gradient for
ionic relaxations. The Z axis is taken perpendicular to the Si(211)
surface, while X and Y axis are along [$\bar{1} 1 1$] and [$0 1
\bar{1}$] respectively. The VASP code ~\cite{vasp} is used for our
calculations.

\begin{figure}
  \includegraphics[width=2.5in]{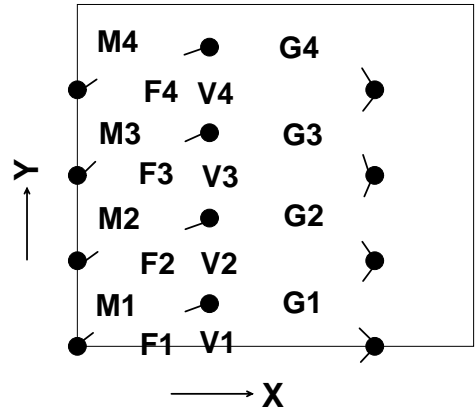}\\
  \caption{Filled circles represent Si atoms of the bulk terminated
  Si(211) surface within a $1\times4$ supercell. First, second, and
  third column of atoms are terrace, trench and edge atoms respectively.
  Dangling bonds of all the atoms are shown. Various kinds of
  symmetric sites are designated as G, V, F, and M sites
  respectively. The numerals 1, 2, 3, and 4 indicate identical
  sites displaced along the Y direction in the super-cell. For
  example, G1, G2, G3 and G4 (Groove sites) are identical sites displaced
  by 3.84 \AA~ along the Y direction in the super-cell.}
  \label{fig2}
\end{figure}

\section{Results and Discussions}
\label{sec:results}

The structures and energetics for Al on the Si(211) surface at 1/8,
1/4, 1/2 and beyond 1/2 ML coverages are systematically analyzed
here to examine the formation of Al atomic wires on the Si(211)
substrate. Note that one monolayer corresponds to one atom per Si
atom on the top layer of the ideal bulk terminated surface leading
to $\sim 5.56 \times 10^{14}$ atoms/cm$^{2}$.

\subsection{Structure and energetics of Al at 1/8 ML coverage on Si(211)}

Here we discuss the structure and energetics of Al at 1/8 ML
coverage and hence we need to place one Al atom on the surface of $1
\times 4$ super-cell. At such a low coverage the Al atoms are
practically isolated from each other on the surface. The Si(211)
surface offers various kinds of sites for Al to bind. Based on the
symmetry and available dangling bonds on the surface, we consider
all the probable binding sites which are designated as G, V, M, and
F sites respectively in Fig.~\ref{fig2}. We use numerals 1, 2, 3 and
4 to label identical sites displaced by 3.84 \AA~ along the
$[01\bar{1}]$ direction in the super-cell. For example, G1, G2, G3
and G4 (G sites) are identical sites displaced by 3.84 \AA~ along
the $[01\bar{1}]$ direction in the super-cell. The binding energies
(BE) of the Al atom at different kinds of sites are given in Table
\ref{table1}. The binding energy of the Al atom is defined as BE =
-$[ {\rm E(Al+Si) - E(Si)} - {\rm E_{Al}}]$ where E(Al+Si), E(Si)
and ${\rm E_{Al}}$ are total energy of Al adsorbed super-cell, total
energy of the super-cell without Al and atomic energy of Al
respectively.

\begin{figure}
  \includegraphics[width=3.5in]{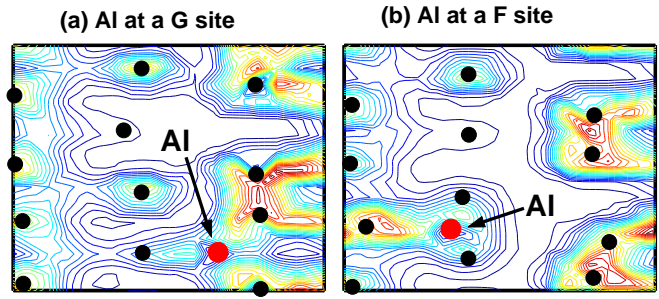}\\
  \caption{Total charge distribution on a plane just above the top
  Si layer at 1/8 ML coverage: (a) when the Al atom is placed at a
  G site and (b) when the Al atom is placed at a F site. In both
  figures, circles in the first, second and third columns are terrace,
  trench and edge Si atoms respectively while the Al atoms are
  indicated by arrows.}
  \label{fig3}
\end{figure}

From Table \ref{table1}, we notice that a G site (BE=4.53 eV) is
most favorable for Al adsorption followed by a F site (BE=4.45 eV),
and a V site (BE=4.31 eV) respectively. Binding energy of Al at a V
site is close to that at a M site. In our calculations, we let the
Al atom (except at V sites) relax in all directions so as to reach
the local energy minimum. For V site, the Al atom is allowed to
relax both along Y and Z directions. It is reasonable that the Al
atom favors to bind at a site where it can satisfy itself by sharing
charges with three neighboring Si atoms. At a G site the Al atom is
surrounded by two edge Si atoms and one trench Si atom. The distance
of the Al atom (at a G site) from the nearest edge Si atoms are
$\sim$ 2.6 \AA~ and that from the nearest trench Si atom is also
$\sim$ 2.6 \AA. On the other hand, the Al atom at the F site also
shares its charge with three Si neighbors and the Al-Si bonds at
those sites are equally strong as that at a G site. For example, the
distances of the Al atom at a F site from its neighboring terrace
and two trench Si atoms are 2.6 \AA~, 2.5 \AA~ and 2.5 \AA~
respectively. Furthermore, the charge density plot in
Fig.~\ref{fig3} clearly shows that the Al atom at both the G and F
sites makes strong bonds with three neighboring Si atoms. We also
notice that the edge Si atoms form dimers irrespective of the
position of the Al atom at a G site or a F site. However, as the Al
atom forms bonds with the neighboring Si atoms, it introduces strain
in the underlying lattice. Therefore the net energy gain due to Al
adsorption will also depend on the amount of strain introduced by
the Al atom during the process of forming bonds with the neighboring
Si atoms. From Fig.~3(b) it is evident that one of the edge Si
dimers in the super-cell is distorted due to the strain introduced
by the Al atom adsorbed at a F site. As a whole, a G site turns out
to be most favorable. The net energy gain in this process is the
difference between the binding energy of Al at a G site and its
chemical potential, $\Delta$E = (BE - $|\mu_{Al}|$) $\sim$ 0.85 eV,
where $\mu_{Al} = -3.68$ eV is the bulk chemical potential for
aluminum. However, in practice, the energy gain will be less than
0.85 eV as the bulk chemical potential is the upper bound of
$\mu_{\rm Al}$. In other words, a maximum of $\sim 0.85$ eV is
gained by adsorbing an Al atom at a G site.

Our calculations show that the adsorption of Al atom on the terrace
site is not possible because one needs to supply an energy of $\sim$
1 eV to place an Al atom on the terrace Si atom. We also studied the
energetics for the replacement of terrace, trench and edge Si atoms
by Al atoms. Our results indicate that replacement of trench and
edge Si atoms by Al atoms are energetically not possible, while an
energy of 0.5 eV is gained due to the replacement of a single
terrace Si atom by an Al atom. We note that the chemical potential
for Si ($\mu_{Si}$) is taken to be the bulk value $\sim$ -5.43 eV.
This information will be useful in the later part of our discussion
at higher coverages of Al.

\begin{table}
\caption{Binding energies (BE) for Al at G1, F1, V1 and M1 sites on
the bulk terminated Si(211) at 1/8 ML coverage.} \label{table1}
\begin{ruledtabular}
\begin{tabular}{|c|c|c|c|c|}
  \hline
  {\bf Site:} & {\bf G1} & {\bf F1} & {\bf V1} & {\bf M1} \\
  \hline
  {\bf BE (eV):} & 4.52 & 4.45 & 4.31 & 4.30 \\
  \hline
\end{tabular}
\end{ruledtabular}
\end{table}

\subsection{Structure and energetics of Al at 1/4 ML coverage on Si(211)}

To analyze the structure and energetics of Al on Si(211) at 1/4 ML
coverage we place two  Al atoms on the bulk terminated Si(211)
surface of the $1\times4$ super-cell. Based on the results at 1/8 ML
coverage and on physical grounds, we consider various reasonable
combinations of a couple of sites where Al atoms may prefer to bind.
The composite sites that we consider are G1G3 (one Al atom is placed
at G1 site and the other is placed at G3 site on the surface of the
1$\times$ 4 super-cell), G1F3 (one Al atom is placed at G1 site and
the other is placed at F3 site), G1M3 (one Al atom is placed at G1
site and the other is placed at M3 site), F1F3 (one atom is placed
at F1 site and the other is placed at F3 site) and M1M3 (one atom is
placed at M1 site and the other is placed at M3 site) respectively.
Without doing any calculations and just based on the results at 1/8
ML coverage, one may conclude that the composite site, G1G3 should
be preferable for Al. However, in the presence of two Al atoms the
surface may undergo further reconstructions due to the interaction
with the surface Si atoms and hence, G1G3 may not be the most
favorable composite site. We therefore, perform extensive
calculations for Al adsorption at all the composite sites. The
average binding energies per Al atom are given in Table
\ref{table2}. By examining Table \ref{table2}, it turns out that
G1G3 configuration is the least favorable (BE per Al is $\sim$ 4.2
eV) configuration while the most favorable configuration is M1M3 (BE
per Al is $\sim$ 4.4 eV).

We note that while M sites turned out to be least favorable among
the G, F, V and M sites at 1/8 ML coverage, it becomes most
favorable at 1/4 ML coverage. In other words, the Al atoms move from
the G sites to M sites through lower energy channels (with a barrier
height of $\sim$ 0.3 eV) as the coverage goes from 1/8 ML to 1/4 ML.
The strain induced in the underlying lattice due to the adsorption
of Al atoms are responsible for such happenings. At 1/4 ML coverage,
the surface consists of Al chains lying between terrace and trench
atoms and extending along the [$0 \bar{1} 1$] direction with an
interatomic distance of 7.68 \AA. Therefore the preferable binding
sites for Al and hence the arrangement of surface atoms are highly
coverage dependent.

\begin{figure}
  \includegraphics[width=3.5in]{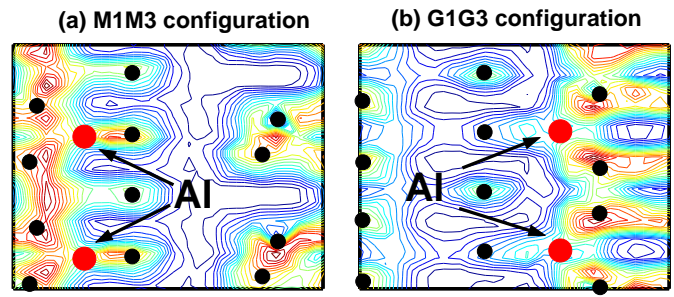}\\
  \caption{The total charge distribution on a plane just below the
  Al layer at 1/4 ML coverage of Al on Si(211): (a) the most favorable
  configuration, M1M3 and (b) the G1G3 configuration. Al atoms are
  indicated by arrows. The small circles represent the terrace, trench
  and edge atoms respectively as one goes from left to right column.}
  \label{fig4}
\end{figure}

The total charge distribution on a plane just below the Al layer for
M1M3 and G1G3 configurations are shown in Fig.~\ref{fig4}. We
observe that the Al atoms in both the configurations form strong
bonds with their neighboring Si atoms. We however note that while
the edge Si atoms in the M1M3 configuration form strong dimers
(dimer length ~ 2.3 \AA) to reduce the total energy of the system,
they are unable to form strong dimers (dimer length ~ 2.9 \AA) in
the G1G3 configuration due to the strain induced by the Al atoms.
This is one of the reasons for favoring M1M3 configuration over the
G1G3 configuration. At this 1/4 ML Al coverage, the average binding
energy per Al atom is slightly lower (by 0.1 eV) than that at 1/8 ML
coverage. We conclude that at 1/4 ML coverage of Al the surface
structure symmetry becomes $1 \times 2$ and the maximum energy gain
per Al atom in the process of adsorbing two Al atoms at the M1 and
M3 sites on the surface of the $1\times4$ super-cell is ~ 0.75 eV.

\begin{table}
\caption{Average Binding energy (BE) per Al atom at 1/4 ML coverage.
A set of two sites combinations on the bulk terminated Si(211) are
considered and they are denoted as G1G3, G1F3, G1M3, F1F3 and M1M3
respectively.} \label{table2}
\begin{ruledtabular}
\begin{tabular}{|c|c|c|c|c|c|}
  \hline
  {\bf Site:} & {\bf G1G3} & {\bf G1F3} & {\bf G1M3} & {\bf F1F3} & {\bf M1M3} \\
  \hline
  {\bf BE (eV):} & 4.22 & 4.26 & 4.33 & 4.37 & 4.43 \\
  \hline
\end{tabular}
\end{ruledtabular}
\end{table}

\subsection{Structure and energetics of Al at 1/2 ML coverage on Si(211)}

\begin{figure}
  \includegraphics[width=3.5in]{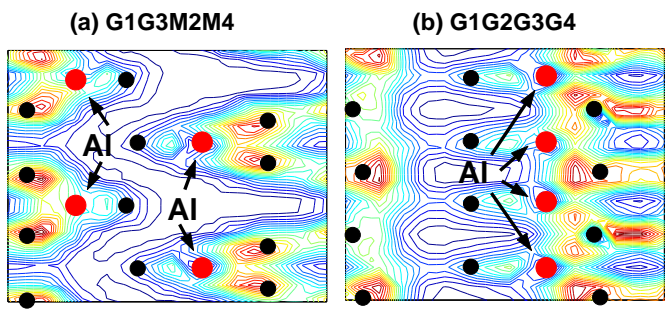}\\
  \caption{The charge distribution just below the Al layer at 1/2
  ML coverage of Al on Si(211): (a) second best favorable configuration,
  G1G3M2M4 and (b) the most favorable configuration, G1G2G3G4. Al
  atoms are indicated by arrows in the figures.}
  \label{fig5}
\end{figure}

The Al coverage is now increased to 1/2 ML {\em i.e.}, four Al atoms
are placed on the surface of the $1\times4$ supercell. Based on
previous results at low coverages and the dangling bonds available
on the surface we consider three plausible combinations of four
sites on the surface. They are designated as M1M2M3M4 (four Al atoms
are placed at M1, M2, M3 and M4 sites available on the surface of
the super-cell), G1G3M2M4 (four Al atoms are placed at G1, G3, M2
and M4 sites available on the surface of the super-cell) and
G1G2G3G4 (four Al atoms are placed at G1, G2, G3 and G4 sites
available on the surface of the super-cell) respectively. The
average binding energy per Al atom for all the configurations are
given in Table \ref{table3}. We note that average energy gain per Al
atom for the configurations G1G3M2M4 (BE $\sim$ 4.3 eV) and G1G2G3G4
(BE $\sim$ 4.4 eV) are very close to each other. The charge density
plot for the G1G3M2M4 configuration is shown in Fig.~\ref{fig5}(a).
We note from Fig.~\ref{fig5}(a) that all the four Al atoms make
bonds with their neighboring Si atoms and the edge Si atoms form
dimers. All the Al atoms remain more or less on the same plane and
saturates all the surface dangling bonds. Therefore, for the
G1G3M2M4 configuration, the highly terraced Si(211) surface becomes
more or less flat decorated by parallel zigzag Al chains separated
by 9.4 \AA~ and extending along the $[01\bar{1}]$ direction.
However, for the most favorable configuration, G1G2G3G4, the terrace
Si atoms remain unsaturated, all the Al atoms remain on the same
plane and make strong bonds with the neighboring trench and edge Si
atoms. Thus, at 1/2 ML Al coverage, the Si(211) surface becomes a
flat surface decorated with straight parallel Al chains separated by
9.4 \AA~ and running along the $[01\bar{1}]$ direction (see charge
density plot in Fig.~\ref{fig5}(b)). The average energy gain per Al
atom in the process of bringing four Al atoms from the source and
putting them at G1,G2,G3, and G4 sites is ~0.7 eV which is
substantial. A comparison of the configurations G1G3M2M4 and
G1G2G3G4 suggests that the complete saturation of surface dangling
bonds does not necessarily lead to the most favorable structure here
and hence the surface strain induced by the adsorbed Al atoms
becomes a significant factor for the most favorable structure. We
therefore conclude that straight Al chains lying between the trench
and edge Si atoms ({\em i.e.}, on the G sites) and extending along
the $[01\bar{1}]$ direction can be formed. We find that the binding
energy per Al atom decreases with increasing coverage.

\begin{table}
\caption{Average Binding energy (BE) per Al atom at 1/2 ML coverage.
A set of four sites combinations on the bulk terminated Si(211) are
considered and they are denoted as M1M2M3M4, G1G3M2M4 and G1G2G3G4
respectively.}
\label{table3}
\begin{ruledtabular}
\begin{tabular}{|c|c|c|c|}
  \hline
  {\bf Site:} & {\bf M1M2M3M4} & {\bf G1G3M2M4} & {\bf G1G2G3G4} \\
  \hline
  {\bf BE (eV):} & 3.88 & 4.30 & 4.37 \\
  \hline
\end{tabular}
\end{ruledtabular}
\end{table}

\begin{figure}
  \includegraphics[width=2.5in]{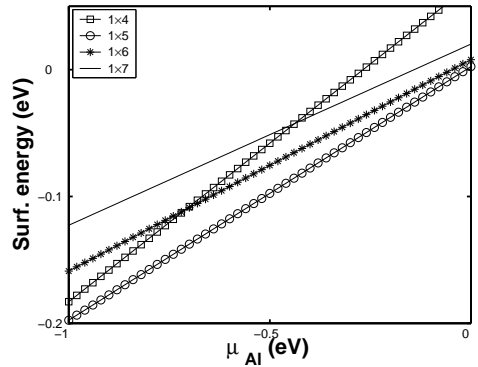}\\
  \caption{The surface energies (eV) for Al chains having vacancies with
  period four ($1\times4$), five ($1\times5$), six ($1\times6$) and
  seven ($1\times7$) relative to the no vacancy case are plotted as a
  function of $\mu_{{\rm Al}}$. Here the surface energy per
  $1\times1$ super-cell surface area ($\sim 36.11 \times 10^{-16} {\rm
  cm}^2$) is considered for comparison among chains with different vacancy
  periods. Note that the chemical potential of Al is given with
  respect to the chemical potential of bulk Al.}
  \label{fig6}
\end{figure}

\subsection{Structure and energetics of Al beyond 1/2 ML coverage
on Si(211)}

The formation of Al chain at 1/2 ML coverage is possible under ideal
conditions because in experiments it is difficult to restrict the
coverage at a predetermined fixed value. Therefore we want to
examine the evolution of surface structure when the Al coverage is
just below 1/2 ML and also when it is somewhat larger, tending
towards 1 ML. A continuous chain lying between the trench and edge
sites produces strain in the system which may be relieved by forming
vacancies along the chain (like Ga \cite{bas1,gonz,bas4}). We
therefore compare the surface energies corresponding to interrupted
Al chains as a function of the chemical potential of Al.
Figure~\ref{fig6} shows the variation of surface energy as a
function of $\mu_{\rm Al}$ for interrupted Al chains (lying on the G
sites) with different vacancy periods. The solid line with squares
correspond to the surface energy for the formation of an Al chain
that is interrupted at every fourth site and resulting surface
pattern becomes $1 \times 4$. Similarly the solid lines with circles
($1 \times 5$), stars ($1 \times 6$) and the solid line with no
symbol ($1 \times 7$) correspond to Al chains with vacancy periods
5,6 and 7 respectively. The surface energies are plotted relative to
the surface energy for the continuous chain (chain without vacancy).
We note that the surface energies for $1 \times 2$ and $1 \times 3$
surface structures are not plotted here because they are very large
compared to those shown in Fig.~\ref{fig6}. We know that the upper
limit of $\mu_{\rm Al}$ is the bulk chemical potential of Al
($\mu^{Bulk}_{\rm Al}$) which is taken as the reference in the
figure. However, the practical value of $\mu_{\rm Al}$ depends on
the experimental conditions. The thermodynamic relation between the
$\mu^{Bulk}_{\rm Al}$ and the practical value of $\mu_{\rm Al}$ is
given by the relation \cite{gonz}
\begin{equation}
\mu_{\rm Al} = \mu^{Bulk}_{\rm Al} - k_B T ln(p_c/p_s)
\end{equation}
where $p_c$ is the Al vapor pressure in the effusion cell and $p_s$
is the Al vapor pressure at the sample and T is the sample
temperature. Therefore, $\mu_{\rm Al}$ is always less than
$\mu^{Bulk}_{\rm Al}$. Experimental conditions can be set up to make
this difference up to $\sim$ 1 eV. We observe from Fig.~\ref{fig6}
that the interrupted chain with a vacancy period of five is
energetically most favorable up to 1 eV below the $\mu^{Bulk}_{\rm
Al}$. However, in the range 0.7 eV $< |\mu_{\rm Al}| <$ 0 eV, the
next favorable chain is the one with six vacancy period. These
interrupted Al chains with vacancy period five and six corresponds
to an Al coverage slightly lower than 1/2 ML. However, to find the
lowest surface energy structure we need to go beyond 1/2 ML
coverage.

\begin{figure}
  \includegraphics[width=2.5in]{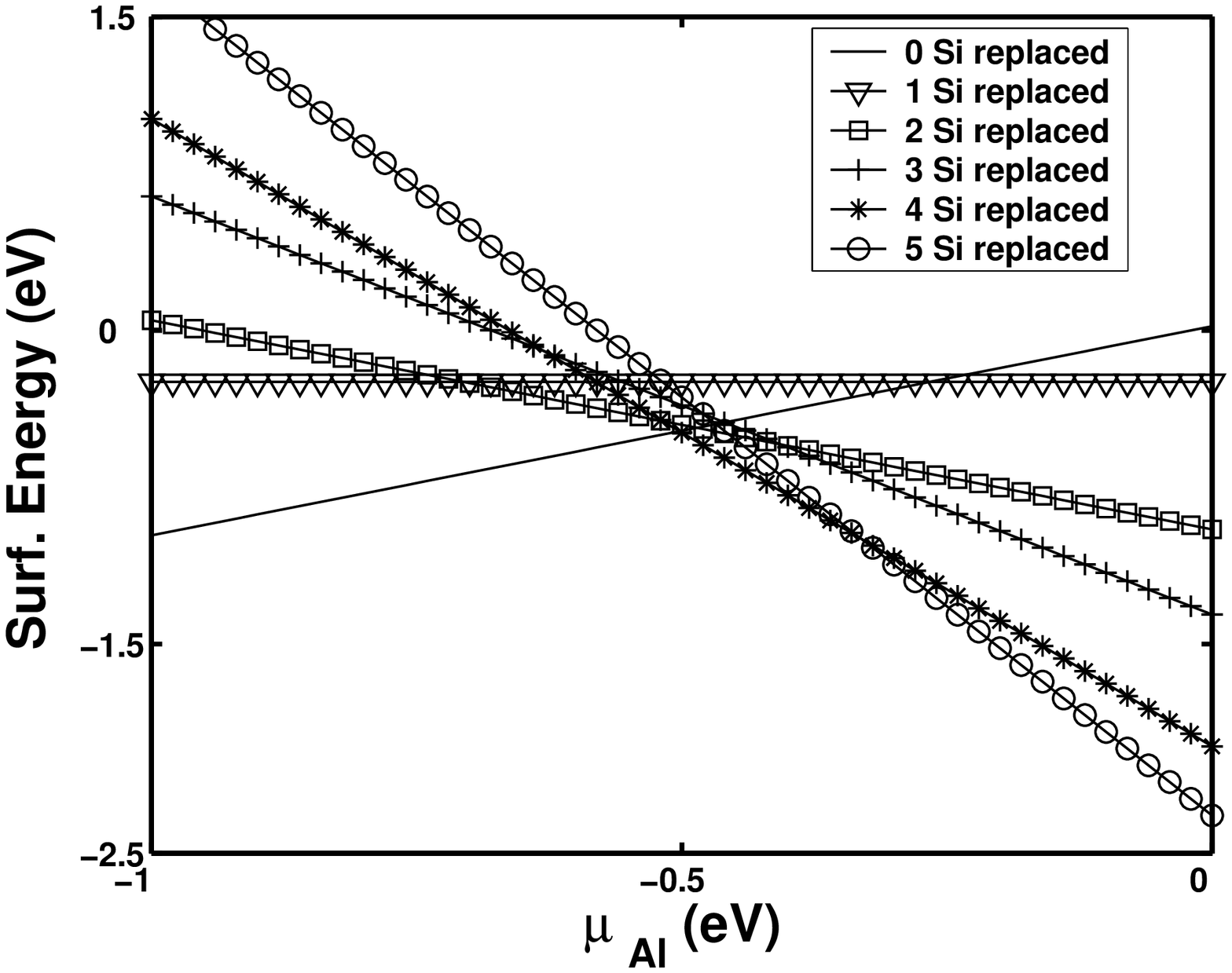}\\
  \caption{The surface energies (eV) as a function of $\mu_{\rm Al}$
  for the situations where zero (solid line), one (solid line with
  inverted triangles), two (solid line with squares), three (solid
  line with plus signs), four (solid line with stars) and five (solid
  line with circles) terrace Si atoms are replaced (in the $1\times5$
  supercell) by Al atoms respectively in the presence of an interrupted
  Al chain with vacancy period five on the groove sites. Surface energies
  relative to that for the no vacancy case is plotted. Note that the
  chemical potential of Al is given with respect to the chemical potential
  of bulk Al.}
  \label{fig7}
\end{figure}

\begin{figure}
  \includegraphics[width=2.5in]{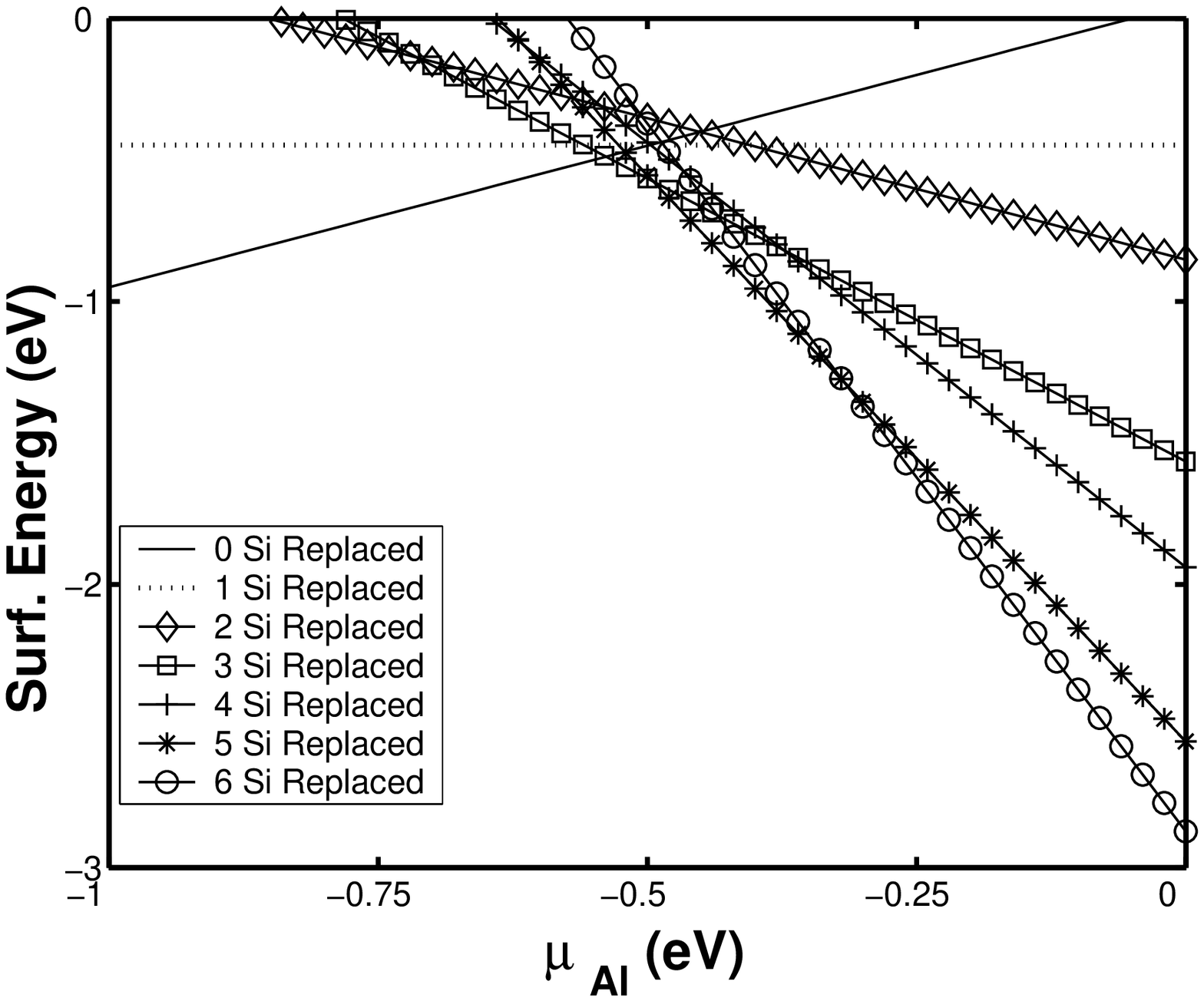}\\
  \caption{The surface energies (eV) as a function of $\mu_{\rm Al}$
  for the situations where zero (solid line), one (dotted line), two
  (solid line with diamonds), three (solid line with squares), four
  (solid line with plus signs), five (solid line with stars) and six
  (solid line with circles) terrace Si atoms are replaced (in the
  $1\times6$ supercell) by Al atoms respectively in the presence of
  an interrupted Al chain with vacancy period six on the groove sites.
  Surface energies relative to that for the no vacancy case is plotted.
  Note that the chemical potential of Al is given with respect to the
  chemical potential of bulk Al.}
  \label{fig8}
\end{figure}

Therefore, we next consider the interrupted chain with vacancy
period five and see how the surface energy changes if we add more Al
atoms to the surface. Although the adsorption of Al atoms at the
terrace sites is energetically not possible, terrace Si atoms may
well be replaced by Al atoms. Figure~ \ref{fig7} shows the variation
of surface energies for replacement of zero, one, two, three, four
and five terrace Si atoms by Al in the presence of an interrupted Al
chain (on the groove sites with vacancy period five) as a function
of $\mu_{\rm Al}$. The solid lines with inverted triangle, square,
plus sign, star, and circle symbols correspond to the situations
where one, two, three, four and five terrace Si atoms are replaced
by Al respectively in the presence of a $1\times5$ interrupted
chain. The solid line corresponds to the no replacement case. We
clearly notice that there are three different regimes for $\mu_{\rm
Al}$ favoring different Al configuration. (i) For, 0 eV $ <
|\mu_{\rm Al}| \lesssim $ 0.3 eV, the replacement of all terrace Si
atoms are favorable in the presence of a $1\times5$ interrupted
chain on the groove sites {\em i.e.}, the Si(211) surface will
consist of one interrupted Al chain with vacancy period five on the
groove sites along with a continuous Al chain on the terrace sites
extending along the [$0 1 \bar{1}$] direction (solid line with
circles). (ii) For 0.3 eV $ < |\mu_{\rm Al}| \lesssim 0.5 $ eV the
Si(211) surface favors an interrupted Al chain on the terrace sites
with period five in the presence of the interrupted Al chain on the
groove sites with a vacancy period five (solid line with stars).
(iii) For $|\mu_{\rm Al}| \gtrsim 0.5$ the surface does favor any
replacement of terrace Si atoms by Al atoms in the presence of the
interrupted Al chain on the groove sites with a vacancy period five
(solid line) {\em i.e.}, the surface consists of only interrupted Al
chains with period five on the groove sites.

We already saw (in Fig.~\ref{fig6}) that the interrupted Al chain on
groove sites with the vacancy period six was the second most
favorable structure in the practical regime of $\mu_{\rm Al}$. It is
therefore desirable to study the possibility of replacement of
terrace Si atoms by Al atoms in the presence of 1$\times$ 6
interrupted chain also. Figure~ \ref{fig8} shows the variation of
surface energies for replacement of zero, one, two, three, four,
five and six terrace Si atoms by Al in the presence of 1$\times$6
interrupted Al chain (on the groove sites) as a function of
$\mu_{\rm Al}$. The dotted line and the solid lines with diamond,
square, plus, star, and circle symbols correspond to the situations
where one, two, three, four, five and six terrace Si atoms are
respectively replaced by Al atoms in the presence of $1\times6$
vacancy chain on the groove sites. The solid line corresponds to the
no replacement case. Similar to the Fig.~\ref{fig7} we again find in
Fig.~\ref{fig8} that there are three different regimes for $\mu_{\rm
Al}$ favoring different Al replacement/no-replacement
configurations. (i) For, 0 eV $ < |\mu_{\rm Al}| \lesssim $ 0.3 eV,
the replacement of all terrace Si atoms are favorable in the
presence of $1\times6$ vacancy chain on the groove sites {\em i.e.},
the Si(211) surface consists of one interrupted Al chain with
vacancy period six on the groove sites along with a continuous Al
chain on the terrace sites extending along the [$0 1 \bar{1}$]
direction (solid line with circles). (ii) For 0.3 eV $ < |\mu_{\rm
Al}| \lesssim 0.5 $ eV the Si(211) surface favors an interrupted Al
chain on the terrace sites with period six in the presence of the
interrupted Al chain on the groove sites with a vacancy period six
(solid line with stars). However, (iii) for $|\mu_{\rm Al}| \gtrsim
0.5$ eV the surface does not favor any replacement of terrace Si
atoms by Al atoms in the presence of the interrupted Al chain on the
groove sites with a vacancy period six(solid line) {\em i.e.}, the
surface consists of only interrupted Al chains with period six along
the groove sites.

\begin{figure}
  \includegraphics[width=2.5in]{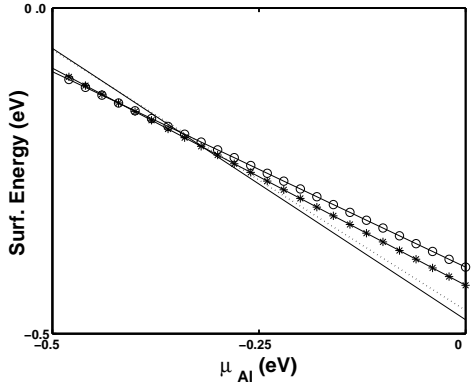}\\
  \caption{The surface energies (eV) per $1\times1$ super-cell surface area
  ($\sim 36.11 \times 10^{-16} {\rm cm}^2$) as a function of $\mu_{\rm Al}$
  for the following surface structures:
  (a) the uninterrupted Al chain on the terrace sites along with an
  interrupted chain on the groove sites with period five vacancy (dotted
  line), (b) the interrupted Al chain on terrace sites with period
  five vacancy along with an interrupted chain on the groove sites with
  period five vacancy (solid line with circles), (c) the uninterrupted
  Al chain on the terrace sites along with an interrupted chain on the
  groove sites with period six vacancy (solid line), and (d) the interrupted
  Al chain on terrace sites with period six vacancy along with an
  interrupted chain on the groove sites with period six vacancy (solid
  line with stars). Surface energies relative to that for the no vacancy
  case is plotted. Note that the chemical potential of Al is given
  with respect to the chemical potential of bulk Al.}
  \label{fig9}
\end{figure}

\begin{figure}
  \includegraphics[width=2.5in]{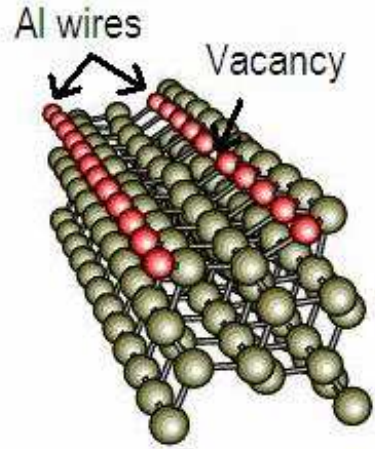}\\
  \caption{The surface consists of two kinds of Al chains (indicated
  by arrows): one of them in uninterrupted and the other one is
  interrupted with a vacancy at a period of six. Both the chains
  are extending along the Y direction. This is the most favorable
  configuration for 0 eV $< |\mu_{\rm Al}| \lesssim$ 0.5 eV.}
  \label{fig10}
\end{figure}

Comparing Figs.~\ref{fig6}, \ref{fig7}, \ref{fig8} we can conclude
that for $|\mu_{\rm Al}| \ge 0.5$ eV the replacement of terrace Si
atoms by Al is energetically unfavorable. The most favorable surface
structure is $1\times5$ where the surface consists of interrupted Al
chains (with vacancy period five) lying between trench and edge Si
atoms and extending along the [$0 1 \bar{1}$] direction. When 0 eV
$< |\mu_{\rm Al}| \lesssim 0.5$ eV, Figs. \ref{fig7} and \ref{fig8}
suggest that the most probable surface structure around one ML Al
coverage can be deduced by comparing the surface energies for four
situations: (a) the uninterrupted Al chain on the terrace sites
along with an interrupted chain on the groove sites with period five
vacancy, (b) the interrupted Al chain on terrace sites with period
five vacancy along with an interrupted chain on the groove sites
with period five vacancy, (c) the uninterrupted Al chain on the
terrace sites along with an interrupted chain on the groove sites
with period six vacancy, and (d) the interrupted Al chain on terrace
sites with period six vacancy along with an interrupted chain on the
groove sites with period six vacancy. These calculations present us
with a rich set of possibilities for creating supported atomic wires
on Si(211). Figure~ \ref{fig9} shows the relative surface energies
for such four surface structures in the range of $|\mu_{\rm Al}|$
between 0 and 0.5 eV. For 0 eV $< |\mu_{\rm Al}| \lesssim 0.3$ eV
the surface structure corresponding to the solid line is most
probable one ($1\times6$) where the surface consists of two kinds of
Al chains extending along the [$0 1 \bar{1}$] direction: one is
continuous and the other is interrupted with a vacancy period six
(see Fig.~\ref{fig10}). However, the surface energy represented by
the dotted line is very close to that represented by the solid line
and therefore the surface structure may also be $1\times5$ where the
surface consists of two kinds of Al chains extending along the [$1 1
\bar{1}$] direction: one is continuous and the other is interrupted
with a vacancy period five. In any case in the range, 0 eV $ <
|\mu_{\rm Al}| \lesssim $ 0.3 eV, an interrupted Al chain with
period five or six appears followed by a continuous Al chain. The
interrupted chain, relieves the strain from the system to enable the
formation of continuous Al chain. In this sense, the interrupted Al
chain assists the formation of the continuous Al chain. In other
words, in the range, 0 eV $ < |\mu_{\rm Al}| \lesssim $ 0.3 eV,
continuous Al chains can be formed \emph{assisted} by the
interrupted Al chains with vacancy period five or six. It is worth
recalling that the formation of continuous Ga chain is not possible
on the Si(211) surface \cite{gonz}. In the range of 0.3 eV $<
|\mu_{\rm Al}| \lesssim$ 0.5 eV, the surface energies for the solid
lines with circles and stars (see Fig.~\ref{fig9}) are lowest and
they are very close to each other and hence we conclude that in this
range of $\mu_{\rm Al}$ uninterrupted Al chains will not be formed,
{\em i.e.} the surface will consist of only interrupted chains and
the surface structure will be either $1\times5$ or $1\times6$. Since
the surface structure and the arrangement of Al atoms on the Si(211)
depends on $\mu_{Al}$ which again depends on the experimental
conditions, the observed surface structure upon Al deposition will
depend on the experimental conditions. In other words, a desired
surface structure among the many (discussed above) will be obtained
by controlling the experimental condition. For example the
uninterrupted Al chain assisted by interrupted Al chain may be
formed if $\mu_{Al}$ is typically maintained within the range of 0.3
eV from the bulk value. Our results indicate that the $1\times6$
surface pattern of Al on the Si(211) observed by Wang et al.
\cite{wang} is one of the most probable structures.

\begin{figure}
  \includegraphics[width=2.5in]{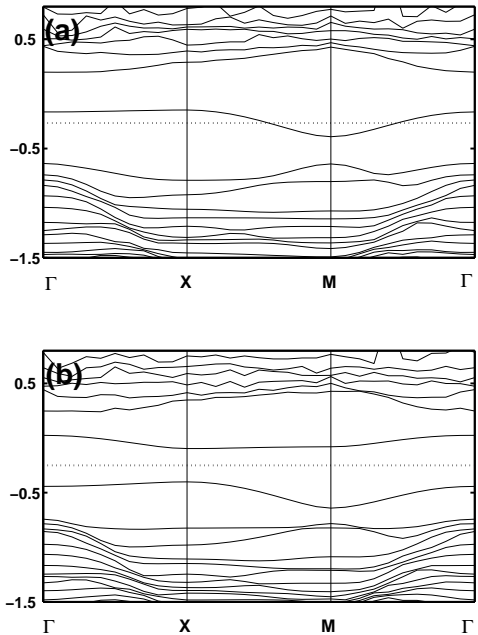}\\
  \caption{Band structures: (a) for the surface consisting of continuous
  Al chains and interrupted Al chains with period $1\times 6$; (b) for the
  surface consisting of only interrupted Al chains with period $1\times 6$.
  The dotted lines indicate the fermi levels.}
  \label{fig10}
\end{figure}

Band structures for two different surface structures are shown in
Fig.~11. In Fig.~11(a), the band structure for the surface
consisting of continuous Al chains and interrupted Al chains with
period $1\times 6$ (most favorable surface structure for 0 eV $<
|\mu_{\rm Al}| \lesssim 0.3$ eV) reveals that the surface is
metallic in nature. On the other hand, Fig.~11(b), indicates that
the band structure for the surface consisting of only interrupted Al
chains with period $1\times 6$ (one of the favorable structures for
0.3 eV $< |\mu_{\rm Al}| \lesssim$ 0.5 eV) is semiconducting in
nature. We therefore conclude that below 1 ML Al coverage the
Si(211) surface may be metallic or semiconducting in nature
depending on the experimental conditions while the Ga covered
Si(211) surface is only semiconducting in nature \cite{gonz}.
However, like Ga/Si(211) an experimental study of Al/Si(211) is
desirable.

\section{summary}
\label{sec:summary}

Energy minimization calculations are performed to study the surface
structures, due to adsorption of Al at various coverages on the
Si(211) surface. Well below 1 ML coverage, we find that though the
average binding energy per Al decreases slowly with increase of the
coverage, the favorable sites change dramatically. For example at
1/8 ML coverage the Al atoms favors to bind at the groove (G) sites
with an average binding energy $\sim$ 4.52 eV and at 1/4 ML coverage
Al atoms prefers to bind at M sites with an average binding energy
$\sim$ 4.43 eV. However, around 1/2 ML coverage Al atoms prefer to
nucleate at the groove sites to form chain structure extending along
the Y[$01\bar{1}$] direction with an interruption by a vacancy at
regular intervals (with vacancy period 5 or 6). As the coverage
increases further, the replacement process comes into play, {\em
i.e.}, the terrace Si atoms are replaced by the Al atoms to form an
uninterrupted/interrupted Al chain along the terrace sites in
addition to the presence of the interrupted Al chain along the
groove sites. A detailed analysis of our results reveals that for
$|\mu_{\rm Al}| > 0.5$ eV, only interrupted Al chain along the
groove sites with a vacancy period of five will be formed and for
0.3 eV $\lesssim |\mu_{\rm Al}| \lesssim$ 0.5 eV interrupted Al
chains with vacancy period five/six will be formed both along the
terrace and groove sites. It is significant to note that for 0 $<
|\mu_{\rm Al}| \lesssim$ 0.3 eV, continuous Al chains along the
terrace sites assisted by the interrupted Al chain (with vacancy
period five/six) along the groove sites can be formed. We note that
the experimentally observed $1\times6$ surface pattern for
Al/Si(211) corresponds to one of the favorable structures presented
here within the 0 eV $< |\mu_{\rm Al}| \lesssim$ 0.5 eV. A novel
finding is that an uninterrupted Al chain may be formed on Si(211)
unlike the case of Ga and In on Si(211). Furthermore, the Al covered
Si(211) surface may be metallic or semiconducting in nature
depending on the experimental conditions, while the Ga covered
Si(211) surface is only semiconducting in nature below 1 ML
coverage. One would expect that since Al (${\rm 3s^23p^1}$) and Ga
(${\rm 4s^24p^1}$) belong to the same column of the periodic table,
they should manifest similar physical and chemical properties. This
is certainly true at a gross level but their behavior at surfaces
can be very different due to differences in lattice constants but
more significantly due to major differences in their cohesive
energies. It should be noted that cohesive energy of Al ($\sim$ 4
eV) is at least $\sim$ 0.5 eV larger (more strongly bound by this
amount) than for Ga. When it comes to binding on surfaces of other
materials, there is a competition between cohesion and bonding and
hence it is not surprising that Al and Ga may stabilize in different
configurations on Si(211). The STM experiments on the Al/Si(211)
system will be helpful in revealing both the continuous and
interrupted Al wire structures.

\end{document}